\documentclass[twocolumn,showpacs,amsmath,amssymb,prl]{revtex4}
\usepackage{graphicx}
\usepackage{dcolumn}
\usepackage{bm}
\usepackage{epsfig}
\usepackage[a4paper]{geometry}
\usepackage[in]{fullpage}
\newcommand{\beq}{\begin{equation}}
\newcommand{\eeq}{\end{equation}}
\newcommand{\beqarray}{\begin{eqnarray}}
\newcommand{\eeqarray}{\end{eqnarray}}

\begin{document}
\title{Tracing Cosmic Accelerators with Decaying Neutrons} 

\author{Reetanjali Moharana}
\email{reetanjali@phy.iitb.ac.in}
\author{Nayantara Gupta}
\email{nayan@phy.iitb.ac.in}
 
\affiliation{Department of Physics, 
Indian Institute of Technology Bombay, Powai, Mumbai 400076, India}
\begin{abstract}
Ultrahigh energy neutrons and pions are likely to be produced in particle interactions inside cosmic ray sources and subsequently decay to neutrinos and other secondary particles [$\pi^{\pm}\rightarrow \mu^{\pm}\nu_{\mu}(\bar \nu_{\mu}),\mu^{\pm}\rightarrow e^{\pm}\bar\nu_{\mu}(\nu_{\mu})\nu_e(\bar\nu_e)$]. In high magnetic fields of the cosmic acceleration sites, the ultrahigh-energy-charged particles may lose energy significantly due to synchrotron radiation before decay. We show that for gamma ray bursts in the internal shock model the flux of very high-energy antineutrinos ($\bar \nu_e$) produced from decaying ultrahigh-energy neutrons can be more than the total neutrino flux produced in pion decay depending on the values of their Lorentz factors, luminosities and variability times. 
  
\end{abstract}

\pacs{98.70.Dk, 98.70.Sa}

\date{\today}
\maketitle
 Pierre Auger \cite{pa1,pa2} and other extensive airshower arrays \cite{hires,agasa,yakutsk} have detected large number of ultrahigh-energy cosmic ray events.
The origin of these cosmic rays are yet to be identified. Some of the ultrahigh-energy cosmic ray events have been correlated with active galactic nuclei (AGN). These events may also come from other sources with similar spatial distributions. Gamma ray bursts (GRBs) have long been speculated as sources of ultrahigh energy cosmic rays \cite{wax1, vietri}. High-energy gamma rays have been detected from many GRBs \cite{fermi}. These gamma rays can be of leptonic or hadronic origin. More observational data may constrain the gamma ray production mechanisms in the near future. As protons can be accelerated to $10^{21}$eV inside GRBs by Fermi mechanism, they are very attractive candidates for multimessenger astronomy.    
 The shock-accelerated ultrarelativistic particles lose energy by interactions and radiations inside their sources and also during propagation. The charged cosmic rays are largely deflected by Galactic and interstellar magnetic fields and it is difficult to trace back their origin. 
   Accelerated protons interact with low energy photons and protons inside their sources. Charged and neutral pions are generated in this way. Charged pions subsequenly decay to high energy secondary neutrinos $p\gamma,pp\rightarrow{\pi^{\pm}}X,\, {\pi^{\pm}} \, \rightarrow \, {\mu^{\pm}}\nu_{\mu}(\bar\nu_{\mu}), \, {\mu^{\pm}} \, \rightarrow \, {e^{\pm}}
\bar\nu_{\mu}(\nu_{\mu})\nu_e(\bar\nu_e)$. Ultrahigh-energy gamma rays are produced from decay of the neutral pions ${\pi}^{0}\rightarrow \gamma \gamma$.
The high-energy gamma rays travelling from high-redshift sources are likely to be absorbed by infrared background and lower-energy photon showers will be produced in cascade interactions. They are found to be useful in probing nearby acceleration sites \cite{murase}. Moreover, ultrahigh-energy photons produced by extremely energetic protons interacting with cosmic microwave background radiation, during propagation through the interstellar medium, can act as messengers of the cosmic accelerators \cite{taylor}.
The ultrahigh-energy secondary neutrinos produced in $pp$ and $p\gamma$ interactions inside the sources are expected to travel to us undeflected by Galactic and extragalactic magnetic fields.
Neutrino telescopes like IceCube \cite{ice}, ANITA \cite{anita} are searching for neutrino signals to complement the gamma-ray detectors. Simultaneous observations by various detectors may reveal many aspects of the same phenomena. 
Neutrino fluxes from GRBs \cite{wax2,nay1,guetta,nay2} and AGN \cite{rac,man,anc1} have been calculated earlier. The neutrino flux from Centaurus A has been calculated \cite{a1,h1,k1} considering $p\gamma$ and $pp$ processes and found to be detectable by IceCube in a few years of operation. 
If the magnetic field inside the sources is very high then the ultrahigh-energy-charged pions and muons lose energy significantly before decaying to secondaries. As a result their secondary neutrino flux is very low.  
 Shock-accelerated protons are expected to produce high energy neutrons in various interactions; these unstable particles subsequently decay to leptons and protons. This channel was previously considered \cite{anc2} to estimate TeV neutrino flux from Cygnus OB2 located about 1.7 kpc away from us. In this paper we discuss that the neutron decay channel of ultrahigh energy neutrino production may become more important than the pion and muon decay channels in case of cosmic accelerators with high internal magnetic fields. In particular, we have considered the neutrino production from GRBs in the internal shock model through photopion and neutron decay, including the effect of synchrotron energy loss by pions and muons in the internal magnetic field of fireball. The values of the Lorentz factor, luminosity, and the variability time of a GRB determine whether this effect will be significant at high energy. In future it would be interesting to study this 
effect in other sources.  The secondary antineutrinos from neutron decay are particularly useful to put upper limit on the energy of the cosmic-ray protons originating from  cosmic acceleration sites of high magnetic fields.                                                 
           
\section{High Energy Neutrinos from Gamma Ray Bursts}
We discuss the photopion and neutron decay channels of high-energy neutrino production inside GRBs.
The high-energy neutrino flux generated in $p\gamma$ interactions through pion
 decay from individual GRBs in the internal shock model has been calculated earlier \cite{nay1,guetta}.
 In \cite{nay2} the authors have derived the neutrino flux from individual GRBs using the isotropic photon energy typically observed by {\it Swift}
 in the energy range of 1 KeV to 10 MeV and folded the neutrino flux with the distribution functions to incorporate the fluctuations in the values of the GRB parameters. The diffuse neutrino flux from all high and low- luminosity GRBs have been calculated in \cite{nay2} considering distribution functions for the GRB parameters. GRBs were assumed to follow star formation rate and the rate of low luminosity GRBs was taken to be 500 times higher than high luminosity GRBs.
In this paper we have calculated the neutrino flux from individual GRBs 
by expressing it in terms of the observed isotropic photon energy as discussed in \cite{nay2} assuming certain values for the luminosity, Lorentz factor and variability time of a GRB. We have not considered the distribution in the values of the GRB parameters which is important in case of the diffuse flux from all GRBs. 
The photon energy spectrum from a GRB can be expressed as a broken-power law with break at $\epsilon_{br}$ in the source rest frame.
\beq
\frac{dn_{\gamma}}{d\epsilon_{\gamma}} 
=A \left\{ \begin{array}{l@{\quad \quad}l}
{\epsilon_{\gamma}}^{-\gamma_1} &
\epsilon_{\gamma}<\epsilon_{br}\\{\epsilon_{br}}^{\gamma_2-\gamma_1}
{\epsilon_{\gamma}}^{-\gamma_2} & \epsilon_{\gamma}>\epsilon_{br}
\end{array}\right.
\eeq

These photons are interacting with shock accelerated protons to produce charged and neutral pions. The charged pions subsequently decay to muons and neutrinos.
Muons also decay to neutrinos. The minimum energy of the protons interacting with photons of energy $\epsilon_{br}$ is  
\beq
E_{pb}=1.3\times10^{7}\Gamma_{300}^2(\epsilon_{br,MeV})^{-1}
{\rm GeV}. 
\eeq
 In $p\gamma$ interactions both $\pi^{0}$ and $\pi^{+}$ can be
produced with equal probabilities. $\pi^{+}$ gets, on average
$20\%$ of the proton energy and if the final-state leptons share the
pion energy equally then each neutrino carries $5\%$ of the initial
proton energy. The first break in energy in the neutrino spectrum,
$E_{\nu b}$ is due to the break in the photon spectrum at $\epsilon_{br}$.
\beq
E_{\nu b}=6.5\times10^5\frac{\Gamma^2_{300}}
{\epsilon_{br,MeV}} {\rm GeV}
\label{Eb1}
\eeq
The fire ball Lorentz factor $\Gamma_{300}= \Gamma/{300}$, photon luminosity $L_{\gamma,51}=L_{\gamma}/(10^{51} ergs \, /sec)$, variability time $t_{v,-3}=t_v/(10^{-3})sec$ are the important parameters in our calculations. The total energy to be emitted by neutrinos of energy $E_{\nu}$ can be expressed as,
\beq
E_{\nu}^2\frac{dN_{\nu}(E_{\nu})}{dE_{\nu}}\approx
\frac{3f_{\pi}}{8}\frac{1}{\kappa}\frac{(1-\epsilon_e-\epsilon_B)}
{\epsilon_e}E_{\gamma}^{iso}
\label{spect}
\eeq
where $E_{\gamma}^{iso}$ is the total isotropic energy of the emitted
gamma-ray photons in the energy range of 1 KeV to 10 MeV; it is
available from observations. $\epsilon_e,\epsilon_B\sim 0.3$ are 
the energy fractions carried by electrons and the magnetic field respectively.
$f_{\pi}$ is the fraction of the energy of a proton going to pion production inside a GRB fireball. The relativistic electrons produce the photons by synchrotron radiation, so four orders of magnitude in photon energy corresponds to two orders of magnitude in the energy of the radiating charged leptons. We have assumed that the electrons 
are radiating all their kinetic energy as observed photon energy $E_{\gamma}^{iso}$. The normalization constant of the power law electron flux is proportional to $E_{\gamma}^{iso}/\kappa$. The differential spectral index of the electron spectrum is assumed to be $-2.5$ which corresponds to $\kappa=1.8$. The total energy carried by the differential flux of neutrinos has been approximately related to the total energy carried by the differential flux of electrons in Eq.(\ref{spect}). The first break [Eq.(\ref{Eb1})] in the neutrino spectrum due to the break energy in the photon spectrum is contained in the expression of $f_{\pi}$ \cite{nay2}. The charged pions and muons lose energy due to synchrotron radiations in high magnetic fields inside the sources before decaying to neutrinos. As a result, the neutrino flux is depleted at very high energy. The neutrino flux from photopion decay in the source rest frame is
\beq
 E_{\nu}^2\frac{dN_{\nu}(E_{\nu})}{dE_{\nu}}\approx\frac{3f_{\pi}}
{14.4}\frac{(1-\epsilon_e-\epsilon_B)}{\epsilon_e}E_{\gamma}^{iso}
 \left\{\begin{array}{l@{\quad \quad}l} 1 & E_{\nu}<E_{\nu}^{s}\\
 (\frac{E_{\nu}}{E_{\nu}^s})^{-2}& E_{\nu}>E_{\nu}^{s} \end{array}
 \right.
\eeq 
Pion-cooling energy is 10 times
higher than muon-cooling energy. We would be overestimating the
total neutrino flux if we use the pion-cooling energy to derive
the second break energy in the neutrino spectrum.
The muon-cooling break $E_{\nu}^s$ can be expressed as a function of
GRB parameters. 
\beq
E_{\nu}^{s}=2.56\times10^{6}\epsilon_e^{1/2}\epsilon_B^{-1/2}
L_{\gamma,51}^{-1/2}\Gamma_{300}^4t_{v,-3}{\rm GeV}
\eeq
   
The fraction of fireball proton energy lost to neutron production is,
\beq
f_n (E_p)= 0.8f_0 \left\{\begin{array}{l@{\quad\quad}l}
\frac{1.34^{\gamma_1-1}}{\gamma_1+1}({E_p}/{E_p}^b)^{\gamma_1-1}& {E_p>E_p^b}\\ 
 \frac{1.34^{\gamma_2-1}}{\gamma_2+1}(E_p/{E_p}^b)^{\gamma_2-1}& {E_{p}< E_{p}^b}
\end{array}
\right.
\label{fn}
\eeq
where $f_0=\frac{0.45L_{\gamma,51}}{\Gamma_{300}^4 \, t_{v,-3}{\epsilon_{br,MeV}}}\frac{1}{\big [\frac{1}{\gamma_2-2}-\frac{1}{\gamma_1-2}\big ]}$, details are discussed in \cite{nay2}. The average value of the fraction of energy transferred from the proton to the neutron is assumed to be $\left< x_{p\rightarrow n} \right>= 0.8$. The neutron spectrum can be expressed in terms of $E_{\gamma}^{iso}$ and the equipartition parameters $\epsilon_e$, $\epsilon_B$.  
\beq
 E_n^2\frac{dN_n}{dE_n}=\frac{f_n}{2\kappa}\frac{1-\epsilon_e-\epsilon_B}{\epsilon_e}{E_{\gamma}^{iso}}
\eeq
$\frac{1}{2}$ corresponds to the assumption that charged and neutral pions are
produced with equal probabilities. The GRB is assumed to be at a redshift of 0.8 in all our calculations. Duration of the burst is assumed to be 5 sec. The break in the photon spectrum
   is assumed to be at $\epsilon_{br}=0.3MeV$ in the source rest frame.
We have assumed the spectral indices of the photon flux $\gamma_1$=1.5 and $\gamma_2=2.2$.
 The ultrahigh energy neutrons decay ($n\,\rightarrow p +e^-+\overline{\nu}_e$) to $\bar\nu_e$. Their rest frame lifetime is $\overline{\tau}_n\,=\,886$ seconds and decay mean free path $10(E_n/EeV)$ kpc. The maximum energy of the neutrons producing $\bar\nu_e$ depends on the distance $D_s$ of the source. If the source is at a large distance then all the neutrons are expected to decay during propagation. The $\bar\nu_e$ flux expected on earth at energy $E_{\bar{\nu}}$ from decaying neutrons ( mass $m_n$) of flux $({dN_n}/{dE_n})$ \cite{anc2} is 
\beq
\frac{dN_{\bar{\nu}}}{dE_{\bar{\nu}}}(E_{\bar{\nu}})=\frac{m_n}{2\,\epsilon_0}\int_{\frac{m_n\,E_{\bar{\nu}}}{2\,\epsilon_0}}^{E_{n,max}}\frac{dE_n}{E_n}\,\frac{dN_n}{dE_n}\left(1-e^{-\frac{D_s\,m_n}{E_n\,\overline{\tau}_n}}\right)\,
\eeq
where $\epsilon_0$ is the mean energy of $\bar\nu_e$ in the neutron rest frame.
 The maximum energy of the shock accelerated protons $E_{p,max}$ in the GRB fireball can be $10^{12}$ GeV. The maximum energy of the secondary neutrons is $E_{n,max}=0.8E_{p,max}$, assuming $80\%$ of the proton's energy goes to a neutron in $p\gamma$ interactions. The maximum energy of the antineutrinos is $Q/m_n=0.001 $ times  the maximum energy of the decaying neutrons, where $Q=m_n-m_p-m_e$.      
The observed antineutrino/neutrino flux on earth is
\beq
\frac{dN_{\nu}^{ob}(E_{\nu}^{ob})}{dE_{\nu}^{ob}}=\frac{dN_{\nu}(E_{\nu})}{dE_{\nu}}\frac{1+z}{4\pi D_s^2} 
\eeq
Figures 1-3 show our calculated neutrino fluxes from photopion decay (dashed line) and neutron decay (solid line) in the observer's frame. 
In Fig.1. the Lorentz factor has been varied from 100 to 500 assuming luminosity $10^{51}$erg/sec and variability time 10ms. For small values of Lorentz factor  and variability time the neutron decay channel can become more important than the photo-pion decay channel at very high energy, as shown in Figs.1 and 3. respectively. Figure 2. shows for high values of luminosity this effect is more significant. The number of neutrino events expected from a GRB in IceCube is very low. The neutron decay channel can give only $10^{-5}$ muon events in $Km^2$ area above 1 TeV energy from a GRB at redshift 0.8, with luminosity $L_{\gamma}=10^{53}$erg/sec, Lorentz factor $\Gamma=300$, duration 5sec, and variablity time $t_v=10$ms. The GRB has been assumed to occur at a zenith angle of $120^{o}$ with respect to an observer at the south pole. The number of total neutrino events contained inside $2km^3$ volume of ice is expected to be $10^{-5}$ from the neutron decay channel and $10^{-4}$ from the pion decay channel. Above $10^{16}$ or $10^{17}$ eV we expect the neutron decay channel to dominate over the pion decay channel depending on the values of the GRB parameters. Neutrino telescopes like ANITA  may be useful for detecting the neutrino fluxes in this energy range.      
\section{Conclusion}
Multimessenger approach to explore the cosmic-ray universe has gained much importance due to the successful operations of the cosmic-ray, gamma-ray and neutrino detectors. In this paper we have discussed that the antineutrino flux produced in decay of ultrahigh-energy neutrons may become more important than the neutrino flux from photopion decay depending on the luminosity , Lorentz boost factor, varibility time of a GRB at very high energy. In future one may consider other cosmic acceleration sites with high magnetic fields to explore the importance of the neutron decay channel at very high energy.     
\section{Acknowledgement}
We are thankful to our referees for valuable comments.
\begin{figure*}[t]
\centerline{\includegraphics[width=3.25in]{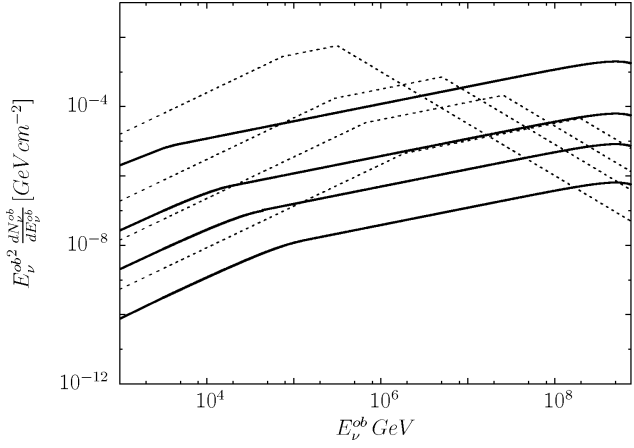}}
\caption {Neutrino flux from pion decay (dashed line) and neutron decay 
(solid line) from GRB with $\Gamma=100$, 200, 300, 500 (top to bottom), $L=10^{51}$ erg/sec, $t_v=10$ ms.}
\label{fig:N1}
\end{figure*}

\begin{figure*}[t]
\centerline{\includegraphics[width=3.25in]{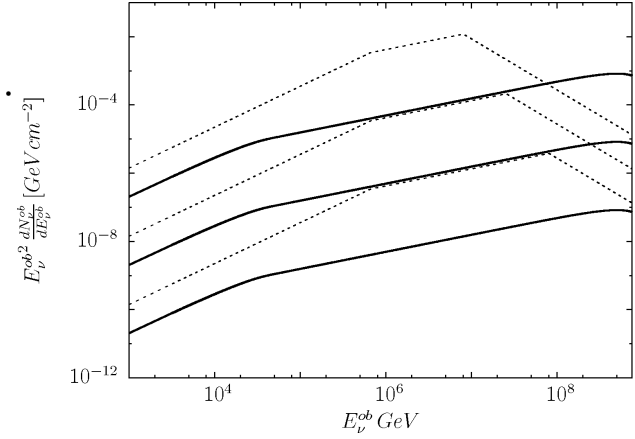}}
\caption{Neutrino flux pion decay (dashed line) and neutron decay (solid line) from GRB with $L=10^{52}$, $10^{51}$, $10^{50}$  erg/sec (top to bottom), $\Gamma= 300$, $t_v=10$ ms.}
\label{fig:N2}
\end{figure*}

\begin{figure*}[t]
\centerline{\includegraphics[width=3.25in]{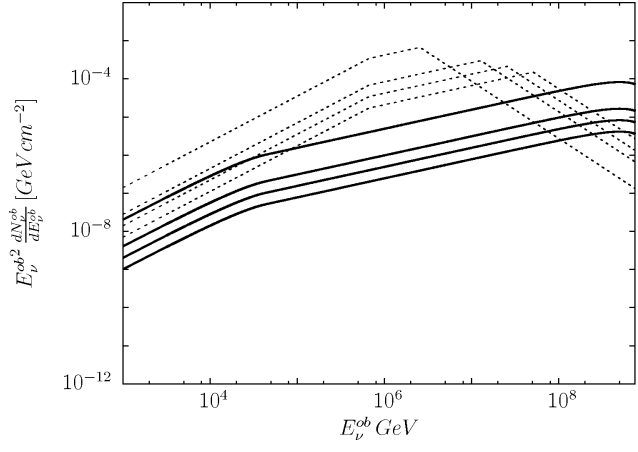}}
\caption{Neutrino flux from pion decay (dashed line) and neutron decay (solid line) from GRB with $t_v=1$, 5, 10, 20 ms (top to bottom), $\Gamma= 300$, $L$= $10^{51}$ erg/sec.}
\label{fig:N3}
\end{figure*}

\end{document}